# IASIS and BigMedilytics: Towards personalized medicine in Europe

Extracting and exploiting clinical value leading with unstructured data by means of Artificial Intelligence and Big Data

Ernestina Menasalvas Ruiz
Centro de Tecnología Biomédica, ETS Ingenieros Informáticos
Universidad Politécnica de Madrid
Pozuelo de Alarcón, Spain
ernestina.menasalvas@upm.es

Alejandro Rodríguez González
Centro de Tecnología Biomédica, ETS Ingenieros Informáticos
Universidad Politécnica de Madrid
Pozuelo de Alarcón, Spain
alejandro.rg@upm.es

Massimiliano Zanin
Centro de Tecnología Biomédica
Universidad Politécnica de Madrid
Pozuelo de Alarcón, Spain
massimiliano.zanin@ctb.upm.es

Consuelo Gonzalo Martin
Centro de Tecnología Biomédica, ETS Ingenieros Informáticos
Universidad Politécnica de Madrid
Pozuelo de Alarcón, Spain
consuelo.gonzalo@upm.es

Juan Manuel Tuñas
Centro de Tecnología Biomédica
Universidad Politécnica de Madrid
Pozuelo de Alarcon, Spain
juan.tunas@ctb.upm.es

Mariano Provencio
Servicio de Oncología Médica
Hospital Universitario Puerta de Hierro
Majadahonda, Spain
mariano.provencio@salud.madrid.org

Fabio Franco
Servicio de Oncología Médica
Hospital Universitario Puerta de Hierro
Majadahonda, Spain
f3franc@gmail.com

Maria Torrente
Servicio de Oncología Médica
Hospital Universitario Puerta de Hierro
Majadahonda, Spain
maria.torrente@salud.madrid.org

Beatriz Nuñez
Servicio de Oncología Médica
Hospital Universitario Puerta de Hierro
Majadahonda, Spain
beangarcia@gmail.com

Virginia Calvo
Servicio de Oncología Médica
Hospital Universitario Puerta de Hierro
Majadahonda, Spain
vircalvo@hotmail.com

*Abstract*—One field of application of Big Data and Artificial Intelligence that is receiving increasing attention is the biomedical domain. The huge volume of data that is customary generated by hospitals and pharmaceutical companies all over the world could potentially enable a plethora of new applications. Yet, due to the complexity of such data, this comes at a high cost. We here review the activities of the research group composed by people of the Universidad Politécnica de Madrid and the Hospital Universitario Puerta de Hierro de Majadahonda, Spain; discuss their activities within two European projects, IASIS and BigMedilytics; and present some initial results.

I. INTRODUCTION

Since it was initially coined, the term Big Data is having an enormous impact in our society. It has gained such importance that governments around the world had to acknowledge its relevance in contexts such as politics, military, law, or management. Accordingly, the European Union followed this trend by creating specific associations and organizations dealing with the impact generated by Big Data and surrounding terms, such as Machine Learning, Artificial Intelligence, etc. If Big Data impacted several fields, the archetype is medicine, as it was soon understood that the incredible amount of routinely generated data could be used for very different purposes. Accordingly, the European Commission has launched several initiatives and calls aimed at funding projects with the objective of studying what are the insights that the extraction, analysis and use of medical data can provide.

The group of Mineria de Datos y Simulación (MIDAS) (Data Mining and Simulation) of the Universidad Politécnica de Madrid (UPM) has followed very closely these movements. With more than 20 years of experience in applying Data Mining (DM) techniques to several fields, the current MIDAS team started several years ago to move its research area to the biomedical domain. Its technical expertise has been complemented by the collaboration with the medical oncology department of the Hospital Universitario Puerta de Hierro de Majadahonda (HUPHM), Madrid, Spain. This has resulted in the involvement in several projects in the context of Big Data and Artificial Intelligence in the medical field, with two of them funded by the European Commission.

This contribution aims at describing the MIDAS / HUPHM team, the expertise of their members, and the techniques by them used. A special focus is given to the two European projects in which they participate, with an analysis of their objectives and characteristics. Some of the results that have been obtained so far in these projects are also presented, as well as other related initiatives.

II. THE TEAM

The meaningful application of Big Data techniques to the biomedical domain requires the convergence of two very different types of expertise: the knowledge of data managing and analysis on one hand and of the medical science on the other. This need buttressed the creation of collaboration between MIDAS and HUPHM, in which each partner contributes as described below.

A. *Universidad Politécnica de Madrid*

The MIDAS group is responsible of tasks related with the extraction of knowledge from medical unstructured data, both in the form of text (electronic health records) and image (CT/PET images). The main people involved are:

- Ernestina Menasalvas Ruiz: Prof. Ernestina Menasalvas is a Full Professor at the "Escuela Técnica Superior de Ingenieros Informáticos" in UPM. She is the principal investigator of the projects described below, and is in charge of the global supervision as well as of tasks in the context of data understanding of text data and of the validation of the results.

- Consuelo Gonzalo Martin: Prof. Consuelo Gonzalo is an Associate Professor at "Escuela Técnica Superior de Ingenieros Informáticos" in UPM. She is the leader of the sub-team inside MIDAS involved in image processing, analysis and understanding. Her main activities include the coordination and supervision of all tasks related to information extraction and structuring, as well as knowledge generation from CT/PET images.

- Alejandro Rodríguez González: Prof. Alejandro Rodríguez is an Associate Professor at the "Escuela Técnica Superior de Ingenieros Informáticos" in UPM and the Principal Investigator of the

Medical Data Analytics Laboratory at Center for Biomedical Technology (CTB). He supervises all efforts related to Natural Language Processing (NLP) tasks.

- Massimiliano Zanin: Dr. Massimiliano Zanin is a post-doctoral researcher at Center for Biomedical Technology at UPM. His main tasks include the supervision of the technical team developing the technical pipeline, and support the work of Prof. Rodríguez and Prof. Menasalvas.

- Juan Manuel Tunas: D. Juan Manuel Tunas is a researcher responsible for the analysis and post-processing of the NLP pipeline results.

### B. *Hospital Universitario Puerta de Hierro-Majadahonda*

Hospital Universitario Puerta de Hierro-Majadahonda is located in Madrid, Spain. This hospital, and more specifically, the medical oncology department, is in charge of providing the definition of the use cases and KPIs regarding the studied pathologies; the required data (electronic health records and images); and, more generally, the expertise necessary for the execution of the projects. The main people involved and their associated areas of responsibility are:

- Mariano Provencio: Medical Oncologist, Chief of Medical Oncology Department at Puerta de Hierro University Hospital, Full Professor, School of Medicine at Autonoma University of Madrid and Scientific Director of the Research Institute at Puerta de Hierro University Hospital. He is the principal investigator of the European projects described below, responsible for the Lung Cancer Pilot.

- Maria Torrente: Medical Doctor and PhD, responsible of international medicine programs in the Medical Oncology Department at Puerta de Hierro University Hospital. Associate Professor, School of Medicine, Francisco de Vitoria University, Madrid. Coordinator of national and international research projects focused on clinical oncology.

- Fabio Franco: Medical oncologist and PhD, within the Lung cancer group in the Medical Oncology Department at Puerta de Hierro University Hospital.

- Virginia Calvo: Medical oncologist and PhD, within the Lung cancer group in the Medical Oncology Department at Puerta de Hierro University Hospital.

- Beatriz Nuñez: Medical oncologist and attending physician in the Medical Oncology Department at Puerta de Hierro University Hospital.

### III. PROJECTS

As previously introduced, the MIDAS / HUPHM group is participating in various projects applying Big Data and Artificial Intelligence to the medical domain. Two of them, both funded by the H2020 programme, are described below.

### A. *IASIS*

Integration and analysis of heterogeneous big data for precision medicine and suggested treatments for different types of patients (IASIS)[1] is a Research and Innovation Action (RIA) funded by European Commission, within its H2020 programme, and under the call "SC1-PM-18-2016 - Big Data supporting Public Health policies[2]". This call targeted projects dealing with the problem of acquiring, managing, sharing, modeling, processing and exploiting huge amount of data within the medical domain, with the goal of developing solutions to support public health authorities. Aligned with the goal defined in the call, IASIS project "*seeks to pave the way for precision medicine approaches by utilizing insights from patient data. It aims to combine information from medical records, imaging databases and genomics data to enable more personalized diagnosis and treatment approaches in two disease areas – lung cancer and Alzheimer's disease*".

---

[1] http://project-iasis.eu/
[2] https://cordis.europa.eu/programme/rcn/700320_en.html

At the current stage of development, IASIS is primarily focusing in an application to lung cancer, being the one on the Alzheimer domain planned for the following months. For this reason, the detailed description of the project objectives are here focused on the lung cancer domain.

IASIS aims to provide answers that can be relevant and effective for the medical practitioners. The main aims regarding lung cancer include:

- Obtaining descriptive and predictive patterns to improve overall survival
- Early detection of relapse and early palliative care initiation, and reducing overtreatments, comparing retrospective datasets with new datasets obtained from our EHR System.
- Implementation of algorithms that reduce drug-drug interactions
- Risk stratification of lung cancer patients (treatment selection based on comorbidity index, family history, risk factors.).

In this context, the main use cases that have been defined in the lung cancer domain include:

- Identifying specific patterns in long surviving lung cancer patients, analysing all the key factors found that may associate to long survival, and compare long-survivors with the rest of the patients, in order to look for specific patterns (natural and family history, treatments, response to treatments, toxicities, comorbidities and molecular mechanisms).
- Search for risk and predictive factors for lung cancer in the study population
- Analyse the effectiveness of tyrosin-kinase inhibitors (TKI) in mutated lung cancer patients (EGFR, ALKt, ROS-1), and look for a possible correlation between toxicities and type/duration of the TKI treatment

The Project is coordinated by National Centre for Scientific Research "Demokritos" (NCSR) in Greece. Beyond UPM and HUPHM, additional partners include: the St. George's Hospital Medical School (UK); Alzheimer's Research (UK); Grupo español de investigación en cáncer de pulmón (Spain); the Centro de Regulación Genómica (CRG) (Spain); the university system of Maryland foundation (USA); and the Gottfried Wilhelm Leibniz Universitaet Hannover (Germany).

### B. BigMedilytics

Big Data for Medical Analytics (BigMedilytics) is an Innovation Action (IA) funded by European Commission, within its H2020 programme, and under the call "*Leadership in enabling and industrial technologies - Information and Communication Technologies (ICT)*"[3]. The projects funded under this call are also known as large-scale pilot projects. The aim of such initiative, in line with the flagship initiative 'Digital Agenda for Europe', includes "*enable[ing] Europe to support, develop and exploit the opportunities brought by ICT progress for the benefits of its citizens, businesses and scientific communities*."

BigMedilytics goals include "*the transformation of Europe's Healthcare sector by using state-of-the-art Big Data technologies to achieve breakthrough productivity in the sector by reducing cost, improving patient outcomes and delivering better access to healthcare facilities simultaneously, covering the entire Healthcare Continuum – from Prevention to Diagnosis, Treatment and Home Care throughout Europe*." BigMedilytics is coordinated by Philips (Netherlands) and the consortium is composed of 35 partners from 11 different countries. Due to the size of the project and the consortium, its organization is divided in several pilots, focusing on the following medical areas/diseases: Comorbidities, kidney disease, diabetes, asthma/COPD, heart failure, prostate cancer, lung cancer, breast cancer, stroke, sepsis, asset management workflows and radiology workflows.

As in the IASIS project, the UPM / HUPHM team is working in the lung cancer pilot. The aim and KPIs defined in this pilot differ from the IASIS project as in IASIS we are more focused in the disease and finding answers to clinical questions that may help us improve our daily clinical practice, while Bigmedilytics is focused in optimizing not only the patient´s management, but also the medical oncology department´s workflow by:

- Increase of early diagnosis: identification of patients at risk of developing lung cancer
- Reducing the cost per patient (reduction of visits to ER, readmissions, reduction of toxicities).
- Reducing toxicity rates specially in complex patients.
- Improving the patients satisfaction: increasing patient´s empowerement and information.

---

[3] https://www.cordis.europa.eu/programme/rcn/664147_en.html

The additional partners involved in the lung cancer pilot of the BigMediliytics project include the National Centre for Scientific Research "Demokritos" (Greece), and the Gottfried Wilhelm Leibniz Universitaet Hannover (Germany).

## IV. DATA SOURCES

The main dataset used in both projects is provided by HUPHM to UPM and includes an anonymized dataset containing data from the Electronic Health Records of 700 lung cancer patients (171.891 clinical notes and 7.021 clinical reports).

### A. IASIS

The IASIS' lung cancer disease area involves the analysis of two different types of unstructured data:

Text: The text in the IASIS project came from the Electronic Health Records (EHR) provided by HUPHM, describing patients diagnosed with lung cancer.

Image: A basic set of images has been provided by HUPHM, for patients with nodules diagnosed as malignity/non-malignity - in a further step, the analysis will be extended to different kinds of malignity. In addition to this, several open access image data bases have been used, including the Lung Image Database Consortium image collection (LIDC-IDRI[4]), NSCLS-Radiomics[5], and LUNA[6].

### B. BigMedilytics

In a similar manner, the lung cancer pilot in BigMedilytics will be executed with similar data (electronic health records provided in IASIS will be also available in BigMedilytics). The main differences in terms of data in BigMedilytics include:

- Image data are not provided: the analysis of the lung cancer information in BigMedilytics is not focused on the analysis of medical images.
- New structured data are provided, based on the specific goals and KPIs of the project:
    1. Oncology calls: The HUPHM provided a set of files containing information about a service for the telephonic attention of cancer patients. This service is not exclusively focused on lung cancer patients. The data recorded contain information about the number of calls performed by each patient and their dates, reasons, and related information.
    2. Oncology app: HUPHM developed a mobile application[7] with the aim of providing cancer patients with information and personalized advices about their disease.

---

[4] https://wiki.cancerimagingarchive.net/display/Public/LIDC-IDRI

[5] https://wiki.cancerimagingarchive.net/display/Public/NSCLC-Radiomics

[6] https://luna.grand-challenge.org

[7] https://play.google.com/store/apps/details?id=org.idiphim.oncoapp

## V. TECHNICAL WORK

As previously explained, both projects (IASIS and BigMedilytics) deal with three types of data:

- Textual information (Electronic Health Records),
- Medical images, and
- Structured data (Call service and mobile app).

For a better understanding of the main technological goals, the methodology associated to each one of these types is described below.

### A. Textual information

The main goal of UPM in this project in terms of the textual information implies:

- To analyze structure free text. UPM is developing a framework called CliKES (Clinical Knowledge Extraction System) based on the Apache UIMA infrastructure. The system is in charge of performing most of the tasks of the classical NLP pipeline, starting with the clinical notes and reports of the patients' EHRs, to end up with a database containing all relevant information. The novelty behind the current work is based on the application of NLP techniques to EHR in Spanish, along with the creation of ad-hoc annotators focused on identifying terms referred to lung cancer domain (including treatments, mutations, etc.).

- To analyze the structured data yielded by the CliKES pipeline, to synthesize useful information for the physicians according to the use cases definition. Specifically, physicians aim at getting relevant information about possible co-occurrences in their lung cancer patients, as well as at trying to find specific correlations between them.

- To provide the results to the IASIS and BigMedilytics consortium, more specifically to the Hannover team, for the creation of knowledge graphs with the information of those patients. Both projects aim at creating a semantic-based version of the data, to allow complex queries with the processed patient data, medical literature knowledge provided by NCSR, and genetic information (in IASIS only, provided by CRG).

### B. Medical images

- Structuring image data. Two types of features have been extracted from the available images: semantic and agnostic features. A python script module has been developed to extract the former ones from CT images in DICOM format. The principal tool used in this module has been PyRadiomics (Van Griethuysen et al., 2017), but also other proprietary libraries. Pre-trained Convolutional Neural Networks (CNN) models have been used for the extraction of agnostic features. The most usual approach of feeding these models is by means of a sliding cube through the 3-D images; yet, this was here completely unviable from a computational point of view. In order to drastically reduce the volume of data to be processed, while minimizing the loss of information, this process has been implemented at a supervoxel level (Gonzalo-Martín et al, 2016).

- In a future phase, the features extracted from images will be used to generate models allowing yielding useful knowledge for physician, such as predicting the survival time of lung cancer patients.

- Finally, provide the results to the IASIS consortium, more specifically to Hannover team, for the creation of knowledge graphs with the information of those patients.

### C. Structured data

Two specific data sources were part of BigMedilytics' lung cancer: HUPHM Oncology call service and OncoApp mobile application. The aim of UPM is to integrate the data generated by these services (which is already in a structured form) to improve the analysis that will be performed; more specifically, to find correlations and patterns in the patients based on their clinical information and their behavior in using these services.

## VI. CONCLUSIONS

As has been shown for all the types of data explained in this paper, a clear relationship is present between the aim of these projects and the application of Artificial Intelligence and Machine Learning techniques. On one hand, both projects have to deal with unstructured data (in image or text form), which require the application of complex techniques and strategies for their handling. In the case of text data, UPM is researching and developing a tool named CliKES, aimed at processing Electronic Health Records in Spanish, something that although has been under development and research by several groups in Spain, is still an on-going task. The nature of the data provided by each hospital and the corresponding processing, the problems associated to the narratives written by each physician, the identification of events, the detection of negation or acronyms in the correct context, the recognition of entities and the appropriate identification of information and the subject that belong to are, among other problems, still open problems in the field of Natural Language Processing, and this despite the large amount of work in the field. Here it is important to emphasize how the machine learning techniques play a very important role in several of the tasks of the NLP pipeline, and how important is to find accurate models to create accurate NLP systems.

Finally, both projects have to deal with structured data. In this context, UPM is mainly working on the application of Data Mining techniques to find important insights and evidences within the data. The amount of data, as well as its diversity (textual, image, call service, mobile application), requires huge efforts in terms of structuring, processing and cleaning. These efforts are done with the objective of having data with enough quality, to subsequently apply the correct data mining techniques and finding evidences based on the use cases, these latter defined by the physicians as well as the associated KPIs.

A comprehensive characterization of lung cancer tumor signatures is critical for a correct diagnosis and optimal treatments. As precision medicine is practiced more widely, one of the main challenges is the integration and analysis of clinical data, opening new opportunities for more accurate diagnosis, more sensitive and frequent disease monitoring and more personalized therapeutic strategies, at the level of the individual.

ACKNOWLEDGMENT

This paper is supported by European Union's Horizon 2020 research and innovation programme under grant agreement No. 727658, project IASIS (Integration and analysis of heterogeneous big data for precision medicine and suggested treatments for different types of patients) and by the European Union's Horizon 2020 innovation programme under grant agreement No. 780495, project BigMedilytics (Big Data for Medical Analytics).